\documentstyle[psfig]{mn}

\def\bm{Block Model}
\def\deltaa{\delta_{\rm b}}
\def\deltat{\delta_{\rm h}}

\def\etal{{\rm et al.\thinspace}}

\def\eg{{\rm e.g.\ }}

\def\ie{{\rm i.e.\ }}

\def\Fig{Fig.}

\def\spose#1{\hbox to 0pt{#1\hss}}
\def\approxlt{\mathrel{\spose{\lower 3pt\hbox{$\sim$}}
        \raise 2.0pt\hbox{$<$}}}
\def\approxgt{\mathrel{\spose{\lower 3pt\hbox{$\sim$}}
        \raise 2.0pt\hbox{$>$}}}
\def\approxpropto{\mathrel{\spose{\lower 3pt\hbox{$\sim$}}
        \raise 2.0pt\hbox{$\propto$}}}
\def\twiddles#1{\spose{\raise 5pt\hbox{$\sim$}}\hbox{$#1$}}
\def\em#1{{\it #1}}

\def\and{, }
\def\ApJ{ApJ}

\def\MN{MNRAS}

\def\Msun{\hbox{$\rm\thinspace M_{\odot}$}}

\title[Merger trees]{Merger trees and the multiplicity function of halos}
\author[D. D. C. Rodrigues and P. A. Thomas]
{D. D. C. Rodrigues and P. A. Thomas\\
Astronomy Centre, Dept.~of Physics and Astronomy, University of Sussex,
Brighton,
BN1 9QH}

\begin{document}

\maketitle
\begin{abstract}
We present a new method for calculating the merger history of matter
halos in hierarchical clustering cosmologies.  The linear density
field is smoothed on a range of scales, these are then ordered in
decreasing density and a merger tree constructed.  The method is
similar in many respects to the block model of Cole \& Kaiser but has
a number of advantages: (i) it retains information about the spatial
correlations between halos, (ii) it uses a series of overlapping grids and
is thereby much better at finding rare, high-mass halos, (iii) it
is not limited to halos whose mass ratios are powers of two, and (iv)
it is based on an actual realization of the density field and so can
be tested against N-body simulations.  The major disadvantages are (i)
the minimum halo mass is eight times the unit cell with a
corresponding loss of dynamic range, and (ii) occasionally the relative
location of halos in the tree does not reflect the correct ordering of
their collapse times, as computed from the mean halo density.
We show that our model exhibits the required scaling behaviour when
tested on power-law spectra of density perturbations, but that it
predicts far more massive halos than does the Press-Schechter
formalism for flat spectra. We suggest reasons why this should be so.
\end{abstract}


\section{Introduction}
\label{secintro}

A basic tenet of modern cosmology is the idea that the present
large-scale structure the Universe originated by the gravitational
growth of small matter inhomogeneities. These initial density
fluctuations are thought to be imprinted in a universe dominated by
collisionless dark matter at very high redshifts. Their distribution
of amplitudes with spatial scale depends ultimately both on the nature
of this collisionless matter and on the physical processes operating
prior to the epoch of recombination.  A family of these generic models
are the moderately-sucessfull hierarchical cosmogonies, which suppose
that the variance of initial fluctuations decreases with scale.  This
means that small structures are the first to collapse and that
galaxies, groups and clusters are formed by the merging of non-linear
objects into larger and larger units.  This merging sequence can be
visualized as a hierarchical tree with the thickness of its branches
reflecting the mass ratio of the objects involved in the merging
(Lacey \& Cole 1993). If we imagine time running from the top of the
tree, the main trunk would represent the final object, while its past
merging history would be represented schematically by the ramification
of this trunk into small branches, representing accretion of small
sub-lumps, and by the splitting into branches of comparable thickness
when merging of sub-clumps of comparable size occurs.

The linear growth of the density field is well-understood, but
collapsed objects, or `dark halos', are highly non-linear
gravitational structures whose dynamical evolution is difficult to
trace. Some progress can be made by the direct numerical integration
of the equations of motion in N-body simulations, but these are
limited in dynamic range and are very time-consuming.  Theoretical
models are usually based on the analytic, top-hat model of Gunn \&
Gott (1972).  Spherical overdensities in a critical density universe
reach a maximum size when their linear overdensity reaches 1.06, then
recollapse and virialize at an overdensity of approximately $\delta_c=1.69$.
Unfortunately real halos are neither uniform nor spherically-symmetric
so that their collapse times scatter about the predicted value.

In cosmology we are seldom interested in the specific nature of one
individual halo, but rather in the statistical properties of the whole
population.  The analytical approach to this problem was pioneered by
Press \& Schechter (1974; hereafter PS).  To estimate what proportion
of the Universe which is contained in structures of mass $M$ at
redshift $z$, the density field is first smoothed with a top-hat
filter of radius $R$, where $M=4/3\pi\overline{\rho}R^3$ and
$\overline{\rho}$ is the mean density of the Universe.  $F(M,z)$ is
then defined to be the fractional volume where the smoothed density
exceeds $\delta_c$.  Assuming a gaussian density field, then
\begin{equation}
F(M,z)={1\over 2}{\rm erfc}\left(\delta_c\over\sqrt{2}\sigma(M,z)\right),
\end{equation}
where $\sigma$ is the root-mean-square fluctuations within the top-hat
filter and ${\rm erfc}$ is the complementary error function.  The key step
was to realize that fluctuations on different
mass-scales are not independent.  In fact, to a first approximation
PS assumed that high-mass halos were entirely made up
of lower-mass ones with no underdense matter mixed in.  Then $F$ must
be regarded as a cumulative mass fraction and it can be differentiated
to obtain the differential one,
\begin{equation}
f(M,z)=-{\partial F\over\partial M}=-{1\over\sqrt{2\pi}}
{\delta_c\over\sigma^2}{\partial\sigma\over\partial M}
e^{-{\delta_c^2/2\sigma^2}}.
\end{equation}
The main drawback of this approach is that, because of the above
assumption of crowding together of low-mass halos into larger ones, it
seems to undercount the number of objects.  As $M\mapsto0$
(and therefore $\sigma\mapsto\infty$) the fraction of the Universe
which exceeds the density threshold tends to one half.  For this reason it
is usual to multiply $f$ by two to reflect the fact that most of the
Universe today is contained in collapsed structures. We call this the
corrected PS prediction.

Extensions of the PS prescription, to calculate explicitly the
integrated merger history of all halos, were first developed by Bower
(1991) and then rederived, using a more mathematically motivated
theory (called the Excursion Set Theory, hereafter EST, Bond \etal
1992), and tested against N-body experiments, by Lacey \& Cole (1993,
1994).  In this formalism the top-hat smoothing radius about a given
point is first set to a very large value and then gradually reduced
until the enclosed overdensity exceeds $\delta_c$ (in hierarchical
cosmologies this will always occur before the radius shrinks to zero).
This gives the largest region which will have collapsed around that
point.  There may be smaller regions which have a larger overdensity
but these merely represent smaller structures which have been subsumed
into the larger one.  By varying the density threshold one can build
up a picture of the collapse and merger-history of the halos: in
essence this paper describes a numerical representation of this process.

Surprisingly perhaps, the EST predicts the same distribution of halo
masses as does the PS theory (but without the need for the extra
factor of two in normalization).  Despite being very idealized in
nature, ignoring both the internal structure and tidal forces, the
derived formulae provide a surprisingly good fit to the N-body results
(Efstathiou \etal 1988, Lacey \& Cole 1994, Gelb \& Bertschinger
1994).  However we have to regard these sucesses with some scepticism,
since the basic hypothesis of the EST works very poorly on a
object-by-object basis (White 1995), the numerical simulations are
still plagued by resolution effects and limited dynamical range, and
the halo statistics are sensitive to the scheme chosen for identifying
halos.  Moreover one should always bear in mind that the PS treatment
is a linear approach to a problem which is fundamentally non-linear in
nature.

The full non-linear evolution of structure is best described by an
N-body simulation.  Moreover, with the introduction of techniques such
as smoothed particle hydrodynamics (SPH), it is possible to
simultaneously follow the evolution of a dissipative, continuous
intergalactic medium.  However, there are several drawbacks to this
approach: N-body simulations are very time-consuming, they have a
limited dynamical range and they are very inflexible when trying to
model the physical processes happening on small scales (with small
numbers of particles).  For example, it is likely that the
interstellar medium in a protogalaxy will contain a mixture of hold
and cold gas as well as stars with a variety of ages and dark matter.
Simulations which can handle such situations are only just
beginning to appear.

Thus it is highly desirable to set up a simple but efficient
Monte-Carlo procedure which mimics the general features of the
hierarchical clustering process and can be used to carry out a large
parameter investigation with little time-consumption. The first model
to be presented in those lines was the \bm\ of Cole \& Kaiser (1989),
used first to study the abundance of clusters and subsequently some
aspects of galaxy formation (Cole 1991, Cole \etal 1994). It starts
with a large cuboidal block, with sides in the
ratio $1:2^{1/3}:2^{2/3}$, and subdivides it into two sub-blocks of
the same shape.  If the initial block has an overdensity $\delta$
(drawn from a gaussian with variance $\sigma(M)$), then the two
sub-blocks will inherit the same overdensity with an extra
perturbation, added to one of them and subtracted from the other,
drawn from a gaussian with variance $\Sigma$, where
$\Sigma^2=\sigma^2(M)-\sigma^2(M/2)$. This quadratic procedure is
applied iteratively to each of the sub-blocks until the imposed mass
resolution is achieved.  The advantage of the method comes from the
fact that the relative position of all sub-blocks is known at all times
so that it is simple to follow the merger history of any halo detected
at any stage of the simulation.

Kauffmann \& White (1993) adopt a different approach which makes use
of the conditional merging probabilities derived by Bower (1991).
Given that a halo has a particular mass at some redshift, then one can
work out the probability distribution for the mass of the halo
(centred on the same point) at some earlier redshift.  By generating
a large number of representative halos, say 100 or more, it is
possible to allocate sub-halos with the correct spectrum of masses.
This method gives a wider mass-spectrum for halos (not restricted to
powers of two) but restricts halo formation to occur at specific
redshifts and is much more complicated to implement than the \bm.

Here we present a new method for following halo evolution which is
much closer in spirit to the N-body simulations without compromising
the simplicity and speed of the above analytical techniques.  It
allows a continuous spectrum of halo masses (above a minimum of 8 unit
cells) and a variable collapse time.  We start with a full realization
of the initial linear density field defined on a cubical lattice.
(This constitutes part of the initial conditions for a cosmological
simulations, which can therefore be used to test our method.)
Secondly we smooth the density field in cubical blocks on a range of
scales, using for each scale of refinement a set of eight displaced
grids.  The blocks are then ordered in decreasing overdensity (\ie
increasing collapse time).  We then run down this list creating a
merger tree for halos.  (The decision whether to merge two sub-halos
together into a larger one is crucial for preventing the growth of
unphysically-large structures.)  As a bonus our
technique retains spatial information about the relative location of
halos (\ie a measure of their separation, not just the merging
topology).

In the next section we describe our merger algorithm in more detail.
Tests on simple power-law spectra of density perturbations are
presented in Section~3, and the relative success, benefits and
disadvantages of our method are contrasted with others in Section~4.


\section{The algorithm}
\label{secmethod}

We begin with a realization of the chosen density field in a periodic
cubical box of side $L\equiv2^l$, where $l$ is a positive integer.  A
standard initial condition
generator is used which populates the box with waves of random phase
and amplitude drawn from a gaussian of mean zero and variance equal to
the chosen input power-spectrum.  Neither the fact that $L$ is a power of two,
nor the periodic boundary conditions are strictly necessary but
are chosen for simplicity.

\begin{figure}
 \centering
 \pssilent
 \psfig{figure=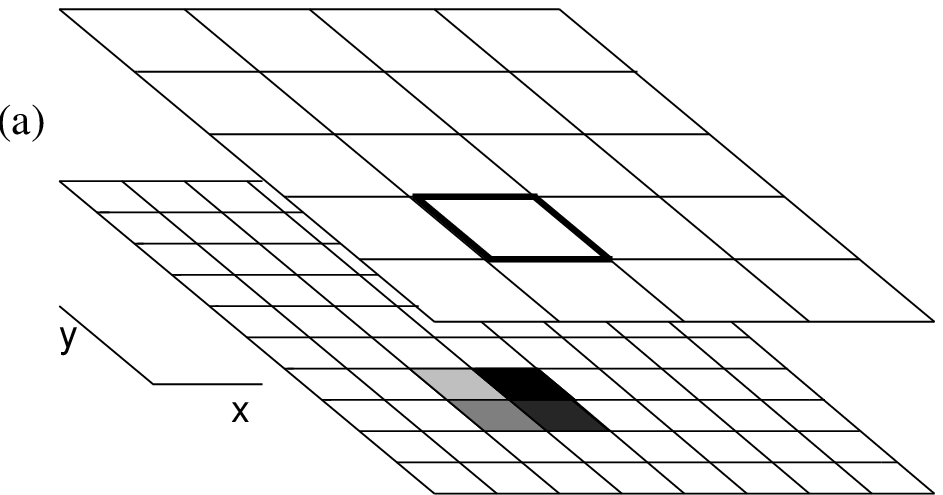,height=4cm}
 \psfig{figure=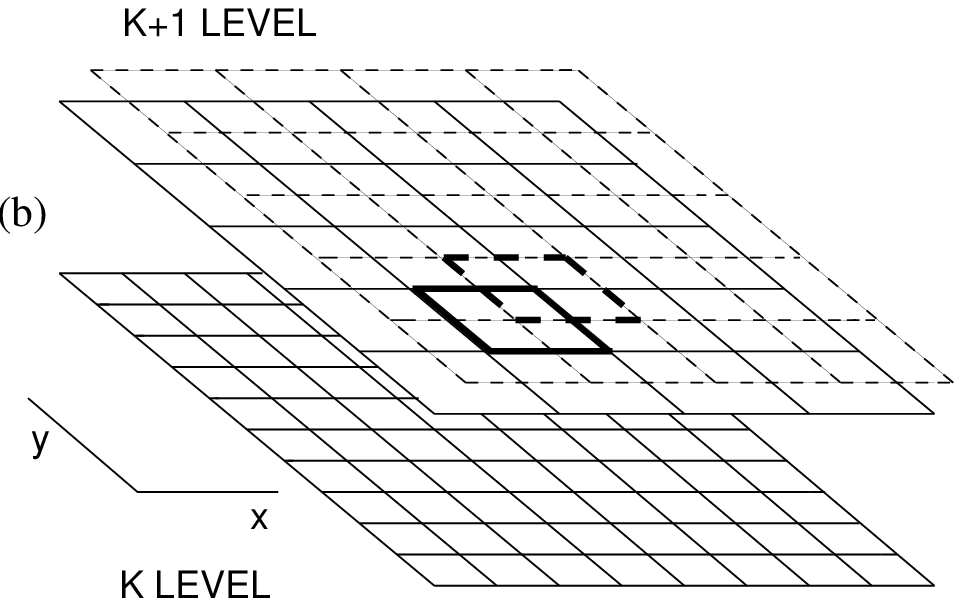,height=5cm}
 \caption{A 2-dimensional representation of the blocking scheme.  (a)
Each block in the upper panel is constructed by averaging the four
cells or blocks beneath it.  (b) This picture shows two sets of
overlapping grids each of which aligns with the same sub-grid from the
previous level of smoothing.}
 \label{figblock}
\end{figure}

Next we average the density fluctuations within cubical \em{blocks}
of side 2, 4, $\ldots$, $L$.  At each smoothing level we use eight
sets of overlapping grids, displaced by half a block-length in each
co-ordinate direction relative to one another (see
\Fig~\ref{figblock}).  This ensures that density peaks will always be
approximately centred within one of the blocks and is a major
advantage over other methods.

The density fluctuations within blocks and base-cells are now ordered
in decreasing density, which is the same order in which they would
collapse as the universe ages (under the na\"{\i}ve assumption that they
all have the same morphology at all times: we will test the accuracy
of this assumption later).

\begin{figure}
\centering
\pssilent
\psfig{figure=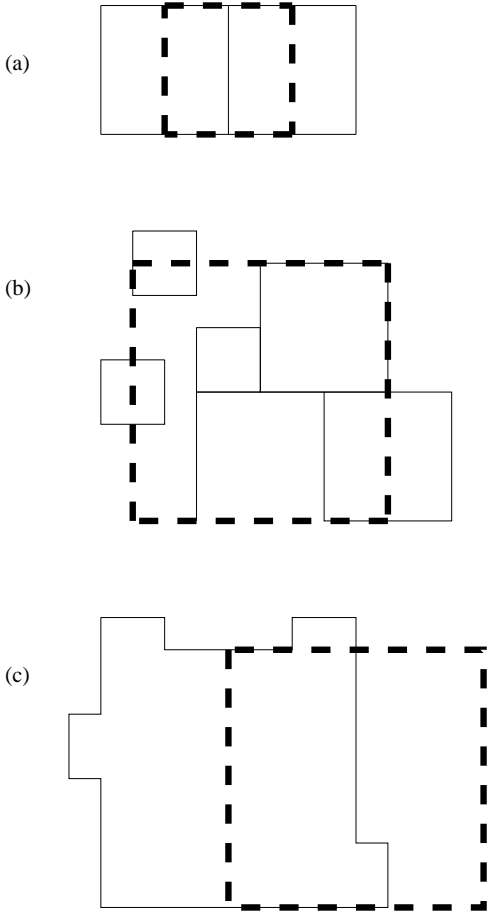,height=12cm}
 \caption{(a) We wish to avoid mergers such as that shown in this
diagram where two pre-existing halos (solid boxes) are linked together
by the collapse of a third block.  This is the reason for the
restriction on merging discussed in the text.  (b) A typical example
of the formation of a new halo by merger of many smaller sub-units.
(c) The growth of a halo by accretion of another block of almost equal
size.}
 \label{fig:halo}
\end{figure}

The final step is to build up a merger tree to express the collapse
history of blocks.  This is a much harder problem than in the simple
\bm\ because the blocks we use are not always
nested inside one another but may overlap.  Our initial guess was to
merge together all collapsed blocks which overlap with one another, but
this leads to very elongated structures which can stretch across a
large fraction of the box.  While these may represent large-scale
pancakes or filaments, they are clearly not the kind of simple
virialized halos which we are trying to identify.  In practice they
would probably break up into smaller objects and so we need to find
some way to limit their growth.  The procedure we use to do this is as
follows:
\begin{itemize}
\item First some terminology. Collapsed regions are known as \em{halos}.
Initially these coincide with the cubical blocks but they need not do
so at later times once overlapping blocks begin to collapse.  The
merger tree consists of a list of cells and sub-halos which
constitute each halo.  (For simplicity each cell, block or halo also
contains a link to its `parent' halo but these are not strictly
required).
\item Initially, there are no collapsed halos. We start at the top of the
ordered list of cells and blocks and run down it in order of
increasing collapse time.
\item Each cell that collapses is given a parent halo, provided that
it has not already been incorporated into some larger structure (this
avoids the cloud-in-cloud problem).
\item For each block that collapses we first obtain a list of all the
halos with which it overlaps and to what extent.  The action to be
taken depends upon this degree of overlap:
\begin{enumerate}
\item Any uncollapsed cells are added to the new halo.  This
represents accretion of intergalactic material.
\item If halos are discovered whose mass is less than that of the
block and at least half of which is contained within the block, then
these are merged as part of the new structure.  This would represent
accretion of existing collapsed objects.
\item If the collapsing block has half or more of its mass contained
in \em{exactly one} pre-existing halo then merge them together as
part of the new structure.  This would represent accretion of the
block by a larger collapsed object.  The restriction to exactly one
pre-existing halo prevents the linking together of adjacent halos
without the collapse of any new matter (see \Fig~\ref{fig:halo}a).  It
is this condition which prevents the growth of long filamentary
structures and limits the axial ratio of halos to be approximately
less than 3:2.
\end{enumerate}
\end{itemize}
Initially the method produces halos of mass 1 and 8 cell units, but as
blocks begin to merge so they produce halos of a wide variety of
shapes and a continuous spectrum of masses.  The two most common
methods of sudden change in halo mass are creation by the merger of
several sub-units (\Fig~\ref{fig:halo}b) or accretion of a new block of
approximately equal mass which overlaps with the halo
(\Fig~\ref{fig:halo}c).  These produce approximately cubic structures,
or triaxial with axial ratios ranging from 3:2 to 1:1 (\Fig~\ref{fig:axes}).
Contrast this with the \bm\ where the halo masses always increase by a
factor of two at each merger event.

\section{Results}
\label{secresult}

\subsection{Self-similarity}
\label{ssecself}

\begin{figure}
\centering
\pssilent
\psfig{figure=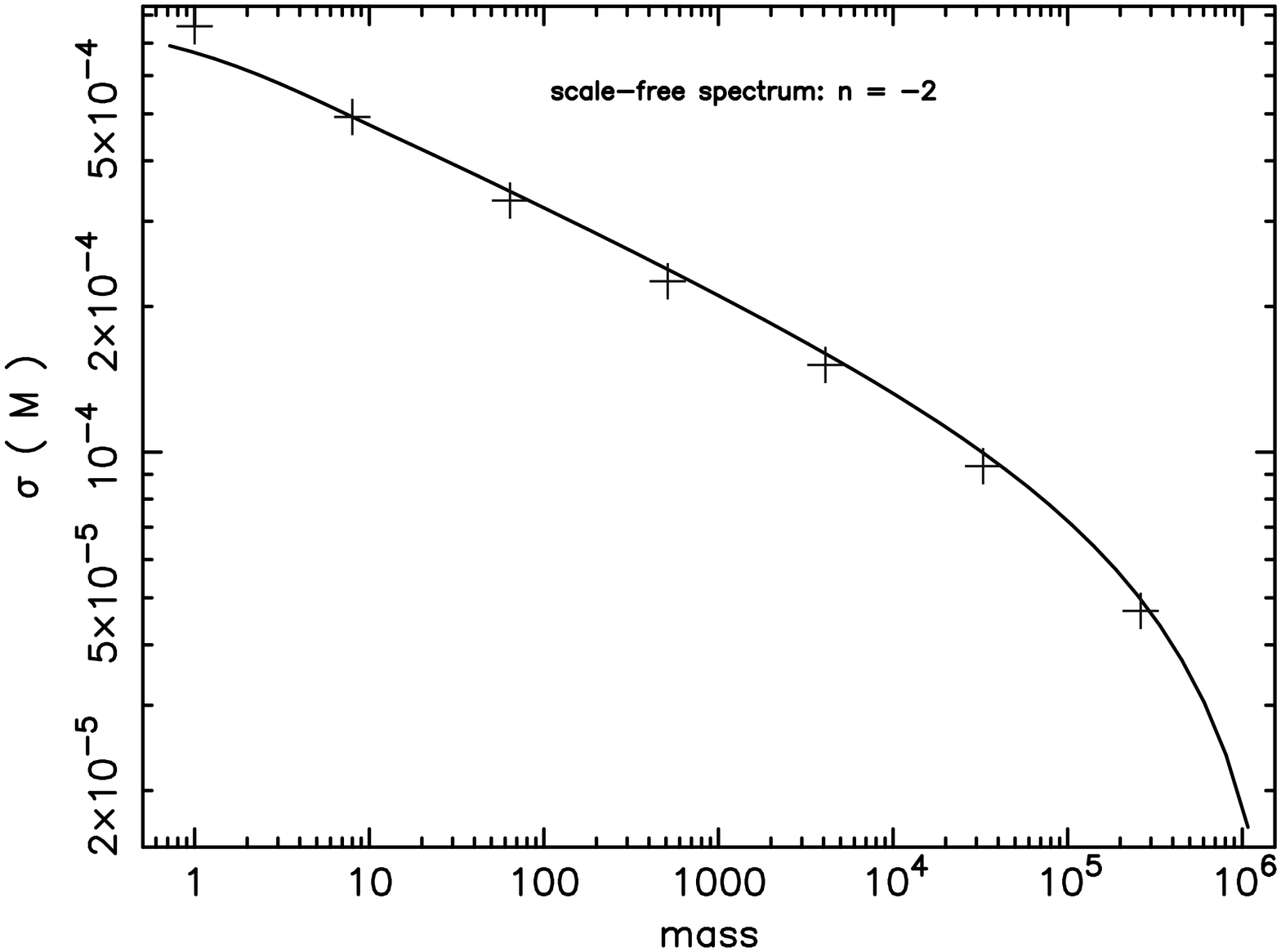,height=6cm,width=8cm,angle=270}
\vspace{0.5cm}
\psfig{figure=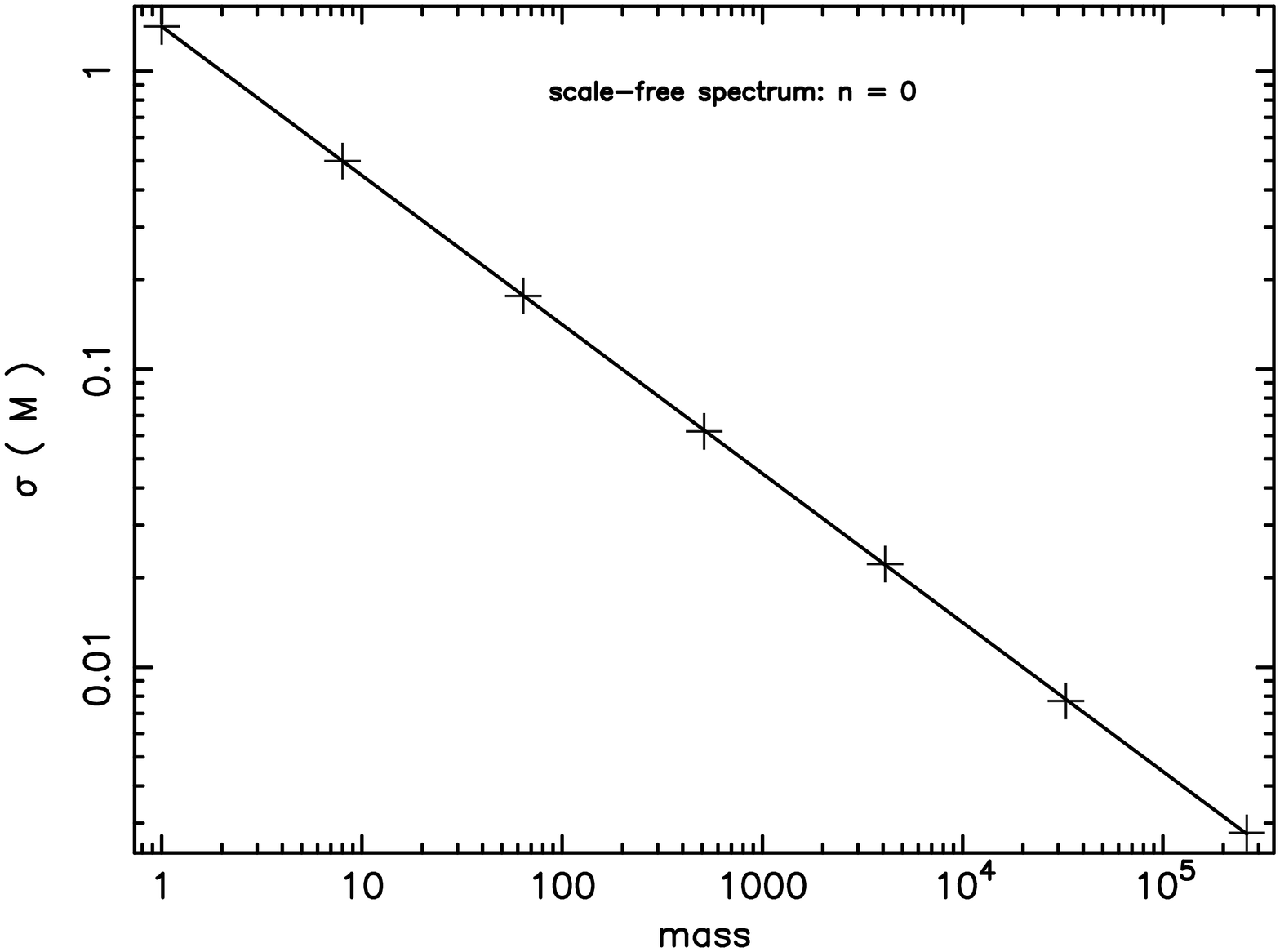,height=6cm,width=8cm,angle=270}
 \caption{The measured root-mean-square power on various mass-scales
for the $128^3$ box: (a) $n=-2$, (b) $n=0$. For $n=-2$ the solid line
is calculated by direct summation of waves inside the box with a
cubical window function. The corresponding curve is normalised to
the second point of the data. The plots are in logarithmic scale.}
 \label{figps}
\end{figure}

We have tested our algorithm on power-law density fluctuation spectra,
which should give self-similar scaling on scales much
smaller than the box-size.  We take a power-law spectrum $P(k)\propto k^n$,
where $n=-2$ or 0 to span the range of solutions expected in the real
Universe.  In an infinite box these would translate to a
root-mean-square density fluctuation spectrum $\sigma(m)\propto
m^{-\alpha}$ where $\alpha=(3+n)/6$.  However, in practice we are
missing a lot of power outside the box and so the decline is steeper
than this at high masses, especially for $n=-2$.  This is illustrated in
\Fig~\ref{figps}a where the spectrum is clearly not a power-law, but
is well-fit by the solid line which shows $\sigma(m)$ calculated by
direct summation of waves inside the box with a window function
associated with a cubical filter. For $n=0$ the effect is not so
severe, so we fit the data with the functional form of $\sigma(m)$ for
an infinite box.  Note that, because we are using a cubical filter,
the normalization is different than it would be for a spherical
top-hat.  This difference is irrelevant for the purposes of this paper
because the normalization we use is arbitrary, however it could be
important if we were to compare our predictions with the results of
N-body simulations.

The results presented here were mostly obtained using boxes of side
$L=128$.
We tried a range of box-sizes, from $L=32$ to 256, to test the effect of
variable resolution on our results.  The code needs about $2L^3$ words
of memory so $L=256$ is the largest practical size on a workstation.
If the merger tree is to be used as the basis of galaxy formation
models, however, then much more storage is required and $L=128$ would
be the largest simulation we can allow for.

\begin{figure}
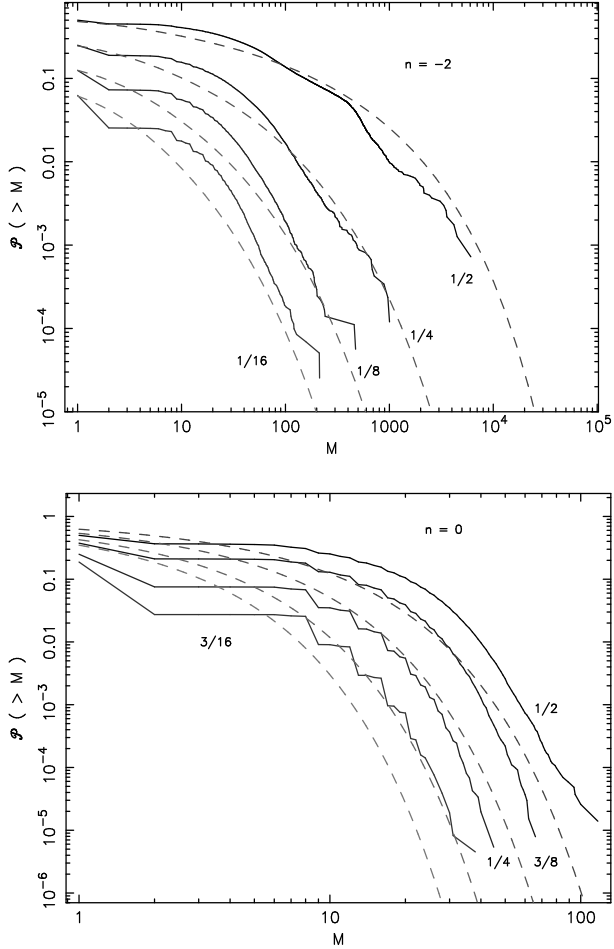

 \centering
 \pssilent
 \psfig{figure=cmf1290.ps.rec,height=6cm,width=8cm,angle=270}
 \vspace{0.5cm}
 \psfig{figure=cmf1288.ps.rec,height=6cm,width=8cm,angle=270}
 \caption{The average cumulative mass function for the four $L=128$
boxes at four different output times (when a certain fraction of the
box is in collapsed regions, as indicated by the figures next to the
curves) for (a) $n=-2$ and (b) $n=0$. The dashed lines show the
corresponding Press-Schechter predictions (with extra factor of two).}
 \label{fig:cummf}
\end{figure}

\begin{figure}
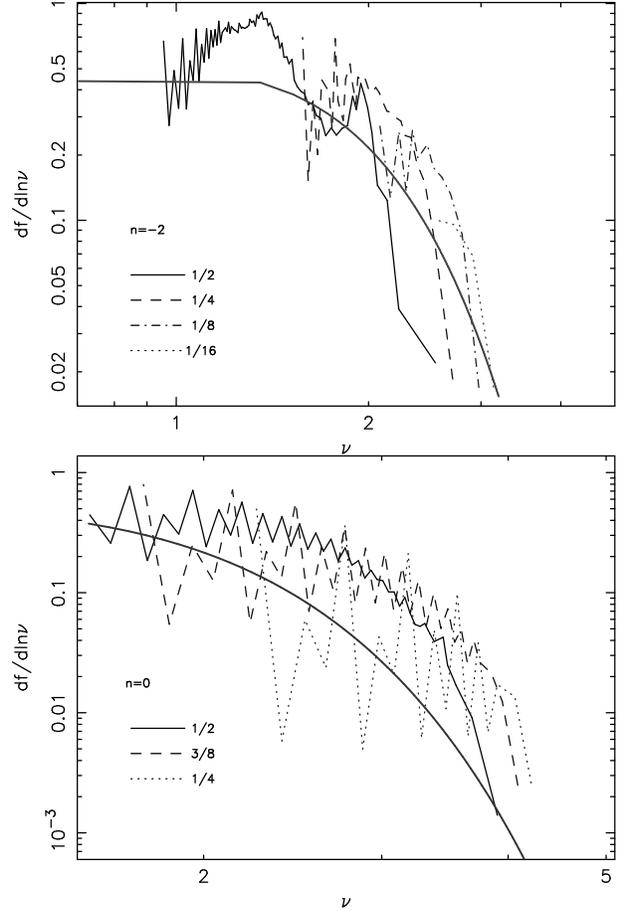

 \centering
 \pssilent
 \psfig{figure=dmf1290.ps.rec,height=6cm,width=8cm,angle=270}
 \psfig{figure=dmf1288.ps.rec,height=6cm,width=8cm,angle=270}
 \caption{The differential mass function, $\partial
 f/\partial\ln \nu$, for the $L=128$ box at the times corresponding to
 the indicated fractions of
 the box contained in collapsed regions: (a) $n=-2$, (b)
 $n=0$. The thick dashed line shows the corresponding Press-Schechter
 prediction.}
 \label{figdifmf}
\end{figure}

\Fig~\ref{fig:cummf} shows the cumulative mass function, $F(M,z)$,
for $L=128$, averaged over four realizations.  The
output is shown for four redshifts corresponding to fractions
${1\over16}$, ${1\over8}$, ${1\over4}$ and ${1\over2}$ of the box
contained in collapsed regions for $n=-2$ and fractions ${3\over16}$,
${1\over4}$, ${3\over8}$ and ${1\over2}$ for $n=0$ (these choices were
made simply to get well-spaced curves in the figure: we can
reconstruct the curves at any intermediate time).  The dashed lines
show the corrected Press-Schechter prediction
where $\sigma(m)$ is obtained from fits to the points shown in
\Fig~\ref{figps}.
In both cases the evolution is approximately self-similar.  This can
be seen more clearly in \Fig~\ref{figdifmf} which shows a differential
plot, $-\partial F/\partial\ln\nu$, where $\nu=\delta_c/\sigma(M,z)$
is the ordinate ($\nu=(M/M_*)^{1/2}$ for $n=0$).  Also shown is the
corrected PS prediction,
\begin{equation}
-{\partial F\over\partial\ln\nu}={2\nu\over\sqrt{2\pi}}
e^{-{1\over2}\nu^2}.
\end{equation}
When expressed in this way the functional form of the mass
distribution is absolutely universal, \ie it does not depend on any
parameter of the simulation.

Consider first the $n=0$ case.  Here the differential mass curves seem
to have the same shape as the PS prediction, but with a
higher normalization (alternatively one could say that $\delta_c$
should be reduced slightly so as to shift the predicted curve to the
right).  This is not unexpected and is discussed in
Section~\ref{ssechighm} below.  There is no evidence of a departure
from the PS curve at a mass of 64, corresponding to the size of
smoothing blocks of side 4 (this is in contrast to the $n=-2$ case,
discussed below).  The maximum mass of collapsed halos is quite small,
less than 125 even for the largest box, $L=256$.  Given that the
smallest halos to collapse in our model (apart from isolated
cells) have mass 8, then this gives a very small dynamic range.  We
could force larger objects to form by allowing a larger fraction of
the box to collapse (this would be legitimate if, for example, one
were to regard the whole box as a single collapsed halo) however one
would not then expect the evolution to be self-similar.

The curves for the steeper spectrum, $n=-2$, extend to much higher
masses because the spectrum has much more power on large scales than
for $n=0$.  Here we do see evidence of kinks at the blocking masses of
64, 512 and 4096, especially at the final output time when half the
box has collapsed: there is an excess of halos of slightly higher
mass and a deficit of slightly lower mass than these.  Overall the
spectrum is a reasonable fit to the PS prediction at masses above 100,
but shows and excess between masses of 8 and 100.

\subsection{Properties of halos}

\begin{figure}
 \centering
 \pssilent
\psfig{figure=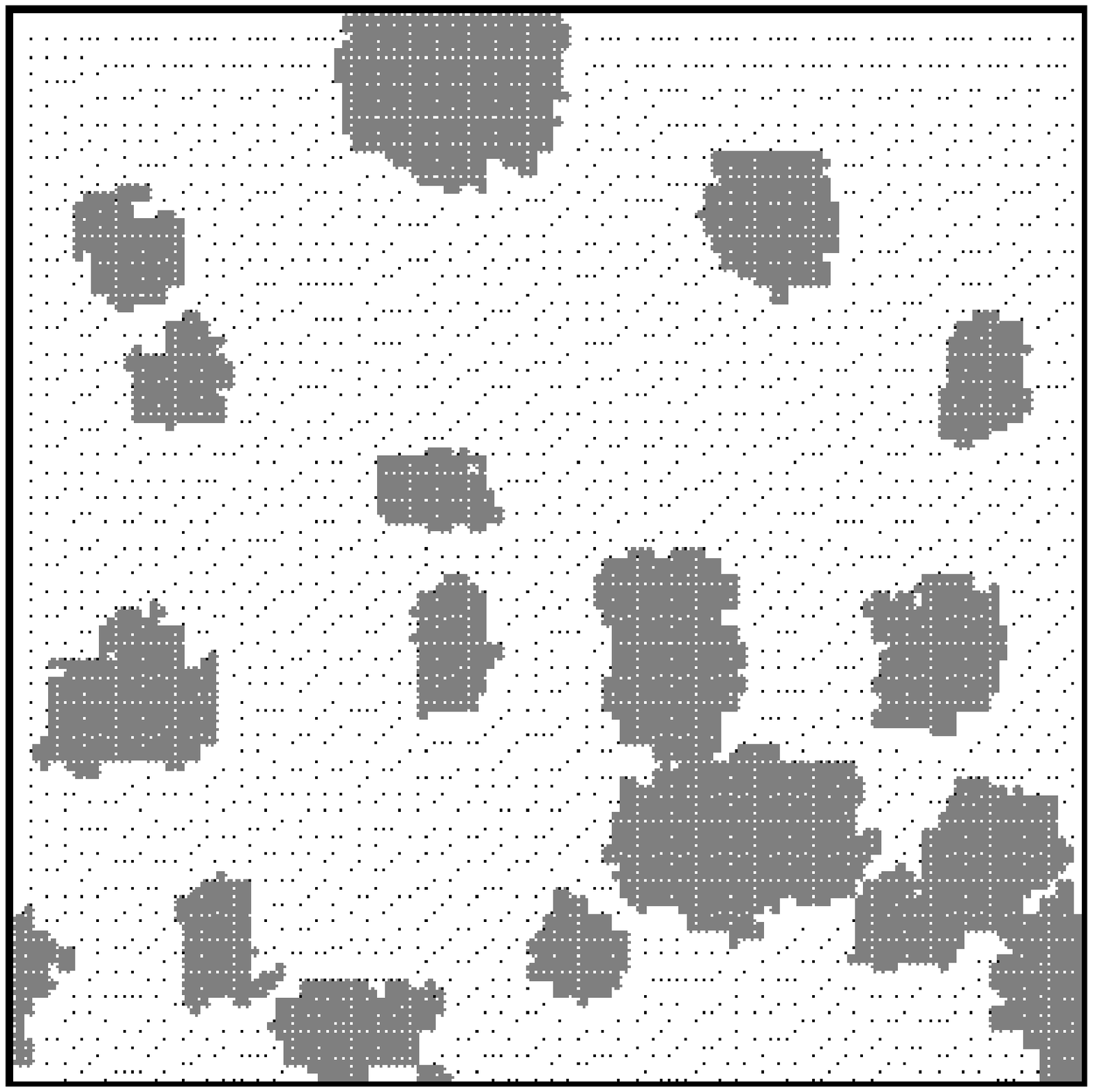,height=7.5cm,width=7.5cm,angle=270}
\vspace{0.5cm}
\psfig{figure=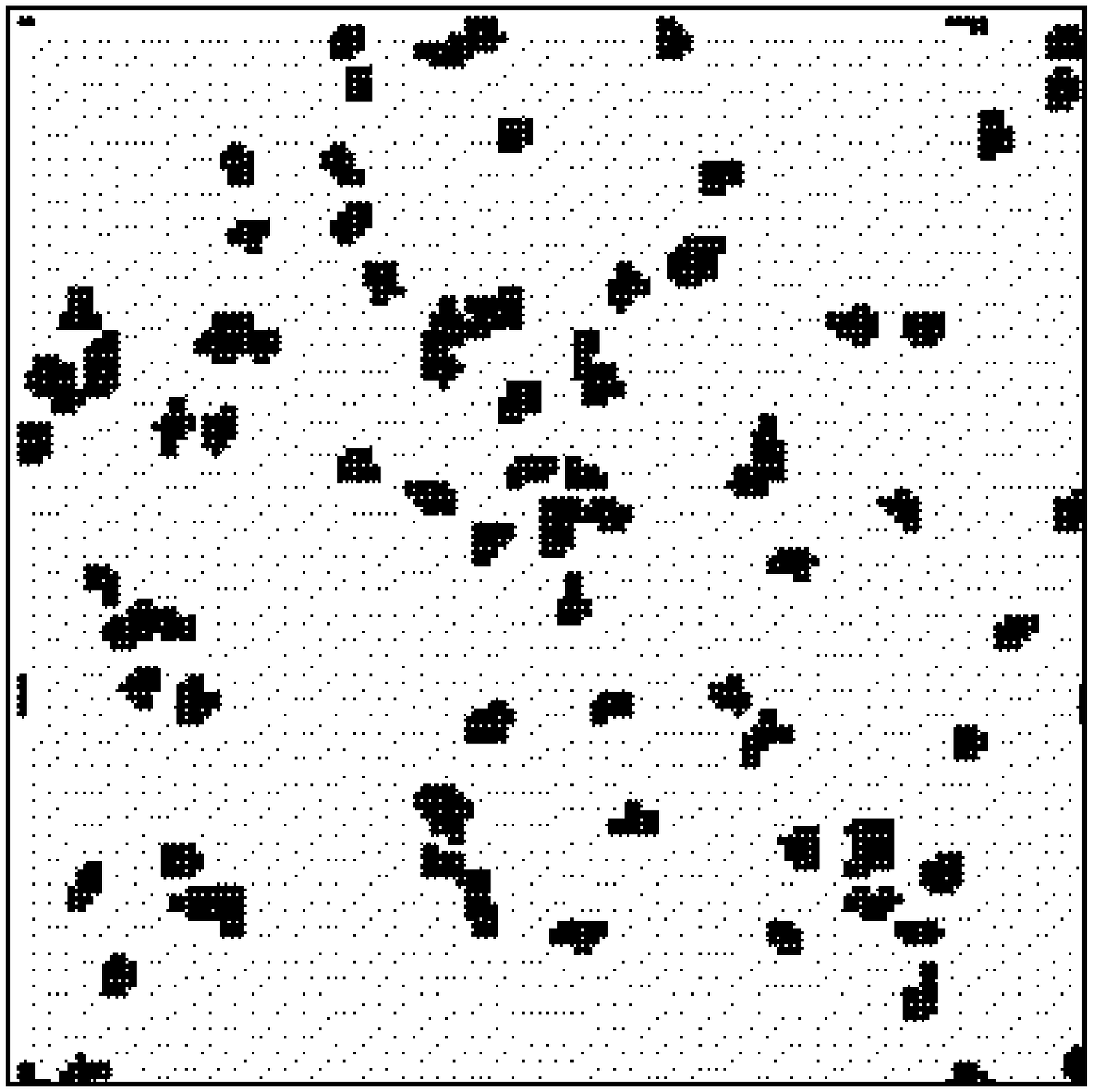,height=7.5cm,width=7.5cm,angle=270}
 \caption{Projections of the distribution of halos taken at the time
when half the box is contained in collapsed structures: (a) $n=-2$,
mass greater than 700; (b) $n=0$, mass greater than 50.}
\label{fig:proj}
\end{figure}

\Fig~\ref{fig:proj} shows a projection of the largest halos in one
$L=128$ box of each spectral type at a time when half the mass has
collapsed into halos.  Many of the irregular shapes which are
visible are due to projection effects.

Our halos tend to exhibit more variety of axial ratios than in
the \bm.  There the relative length of the major- and minor-axes is
fixed all times at approximately 1:1.59, whereas ours start with
more typically 1:1
(for collapse of isolated blocks as in \Fig~\ref{fig:halo}b) or 1:1.5
(for the collapse of overlapping blocks as in \Fig~\ref{fig:halo}c),
developping rapidly to more complex structures with a great variety
of shapes.

\begin{figure*}
\centering
\pssilent
\begin{tabular}{cc}
\psfig{figure=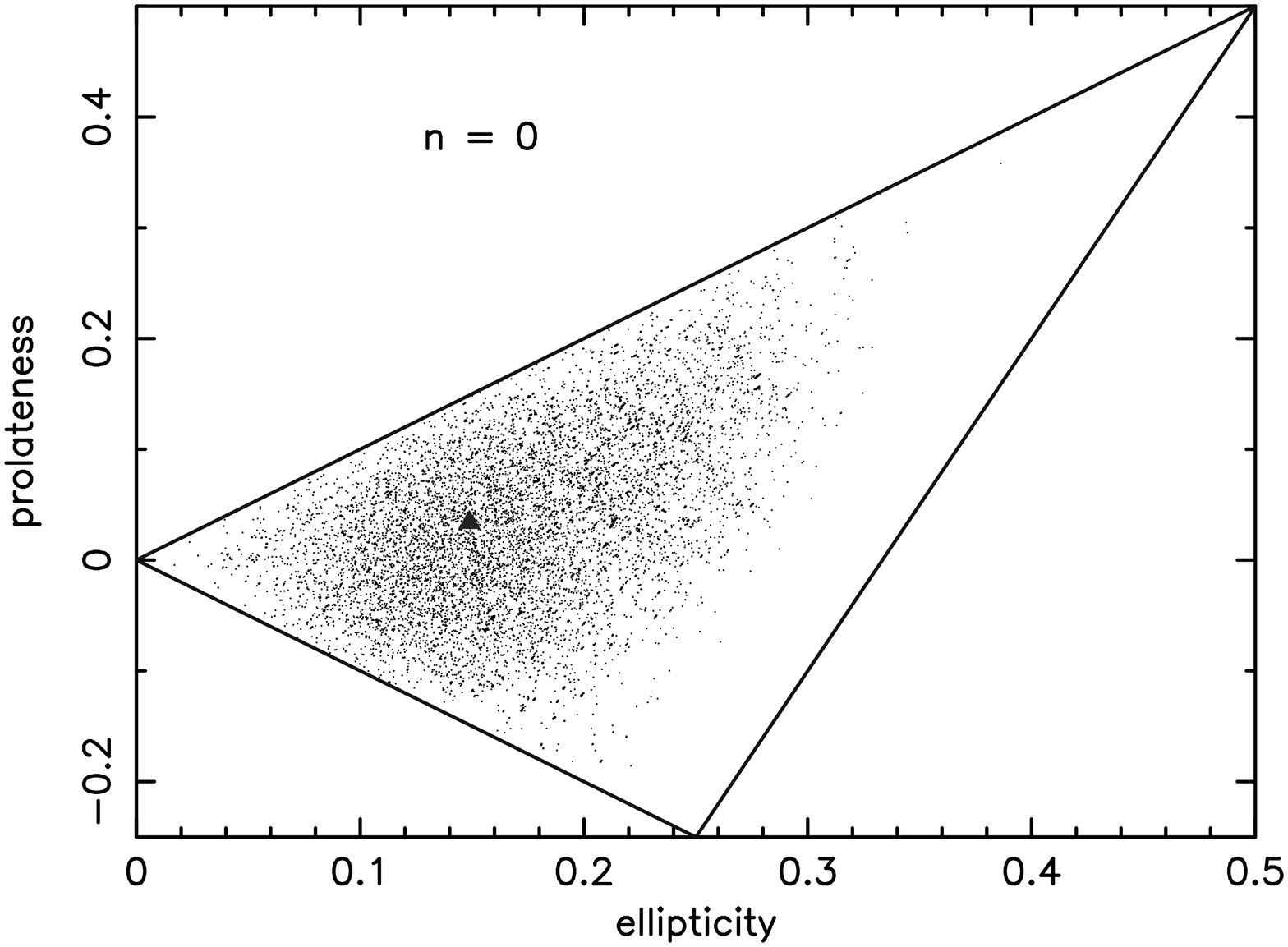,height=7.5cm,width=7.5cm,angle=270} &
\psfig{figure=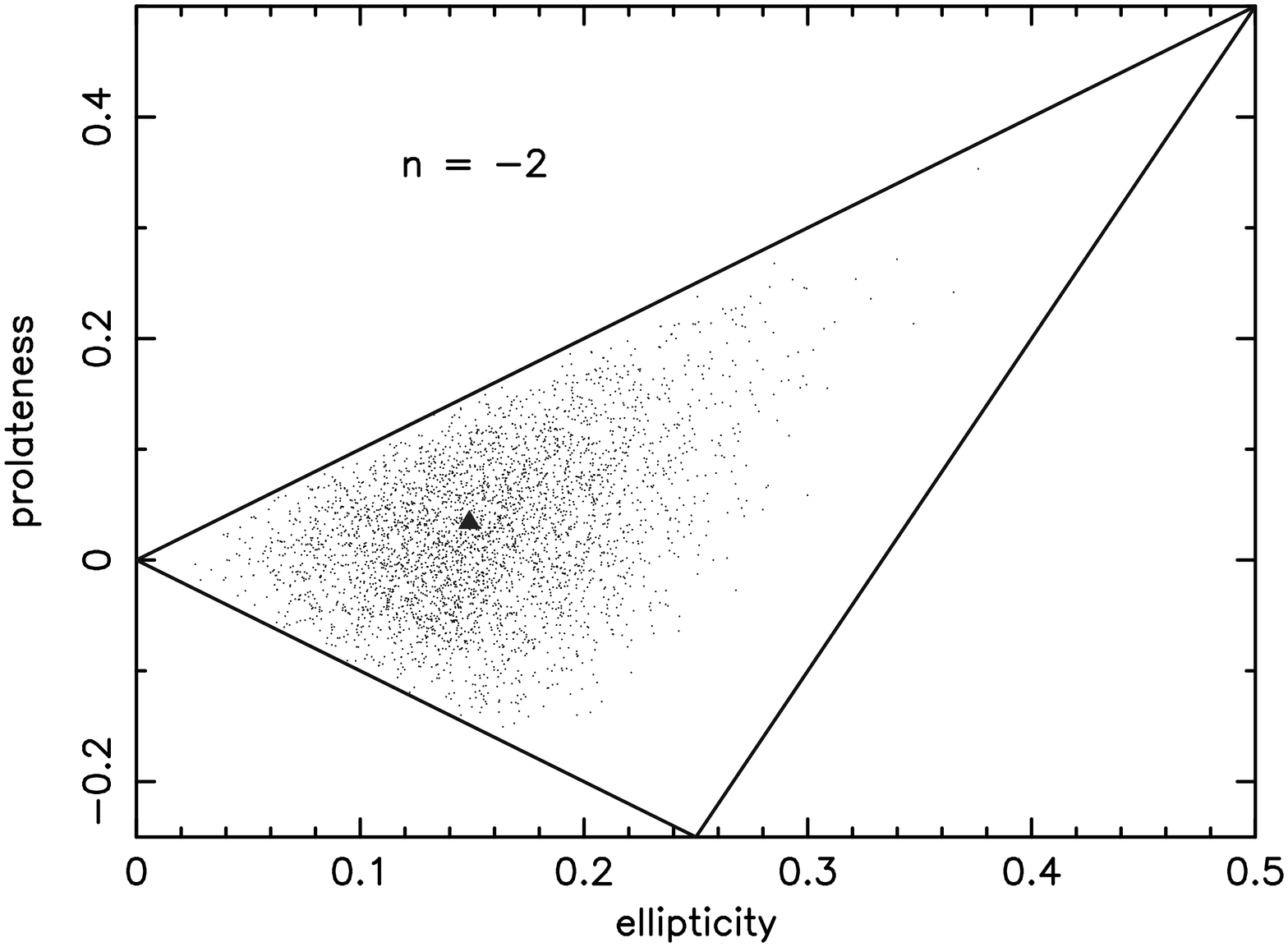,height=7.5cm,width=7.5cm,angle=270} \\
\psfig{figure=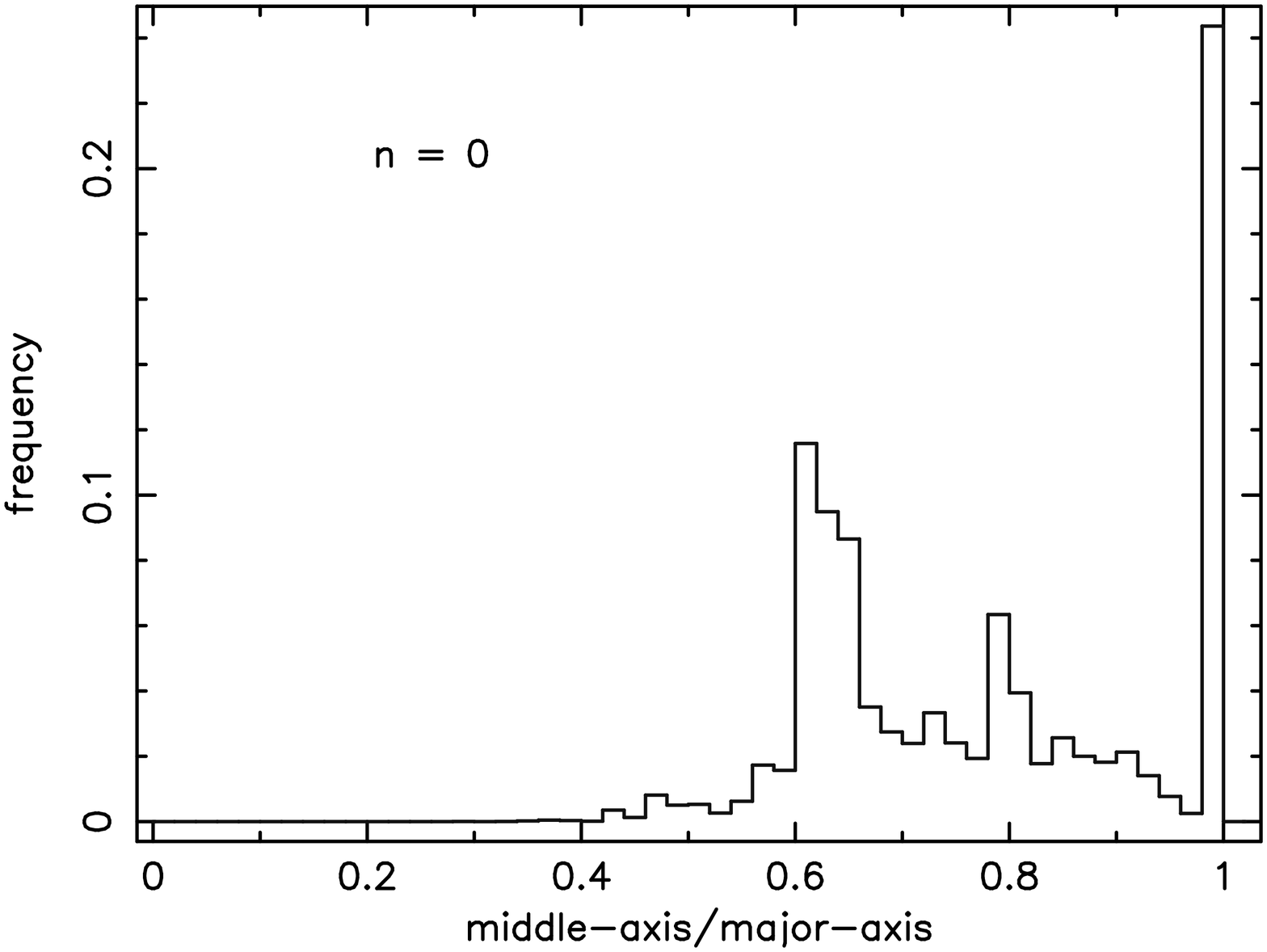,height=7.5cm,width=7.5cm,angle=270} &
\psfig{figure=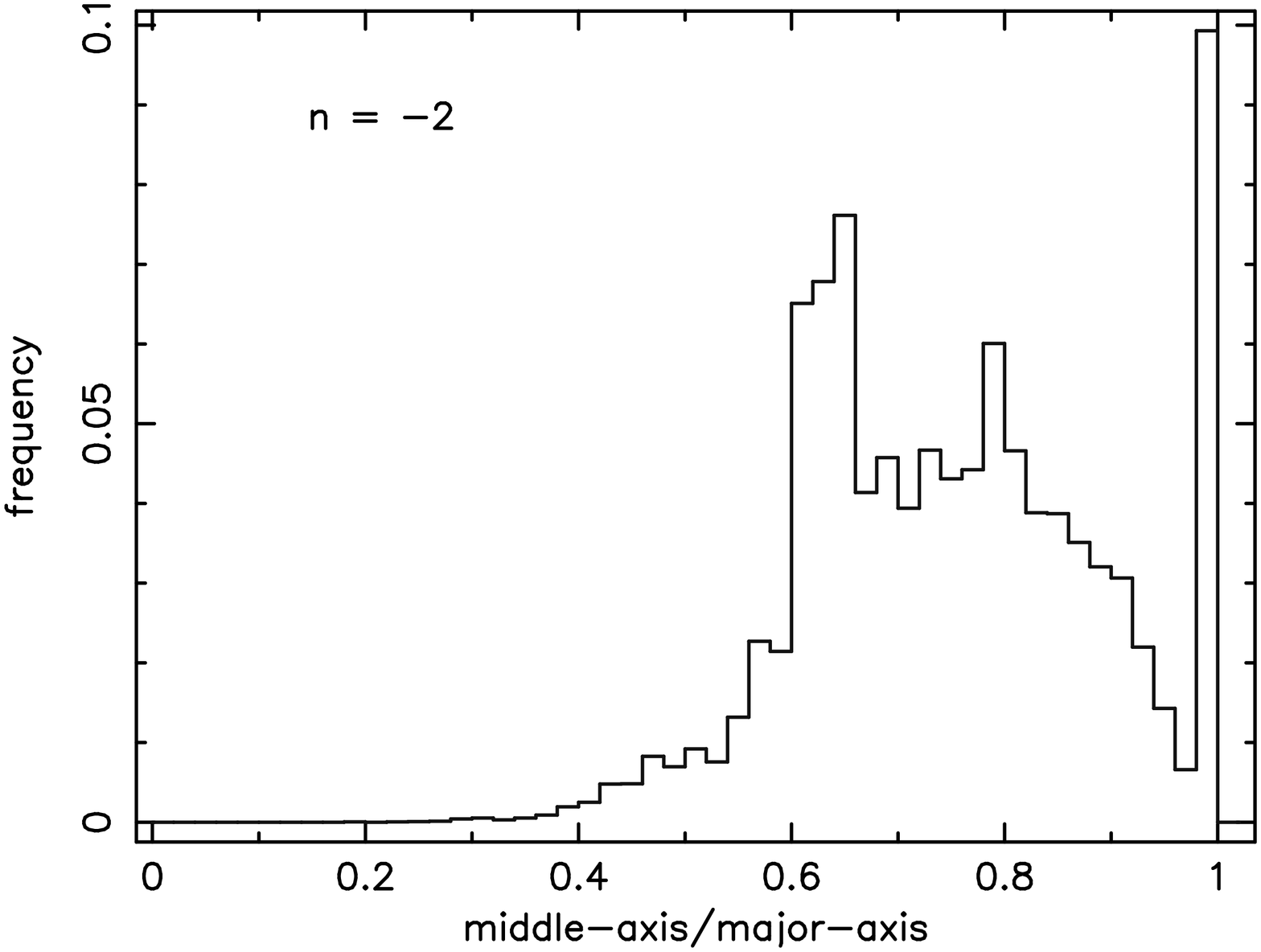,height=7.5cm,width=7.5cm,angle=270} \\
\psfig{figure=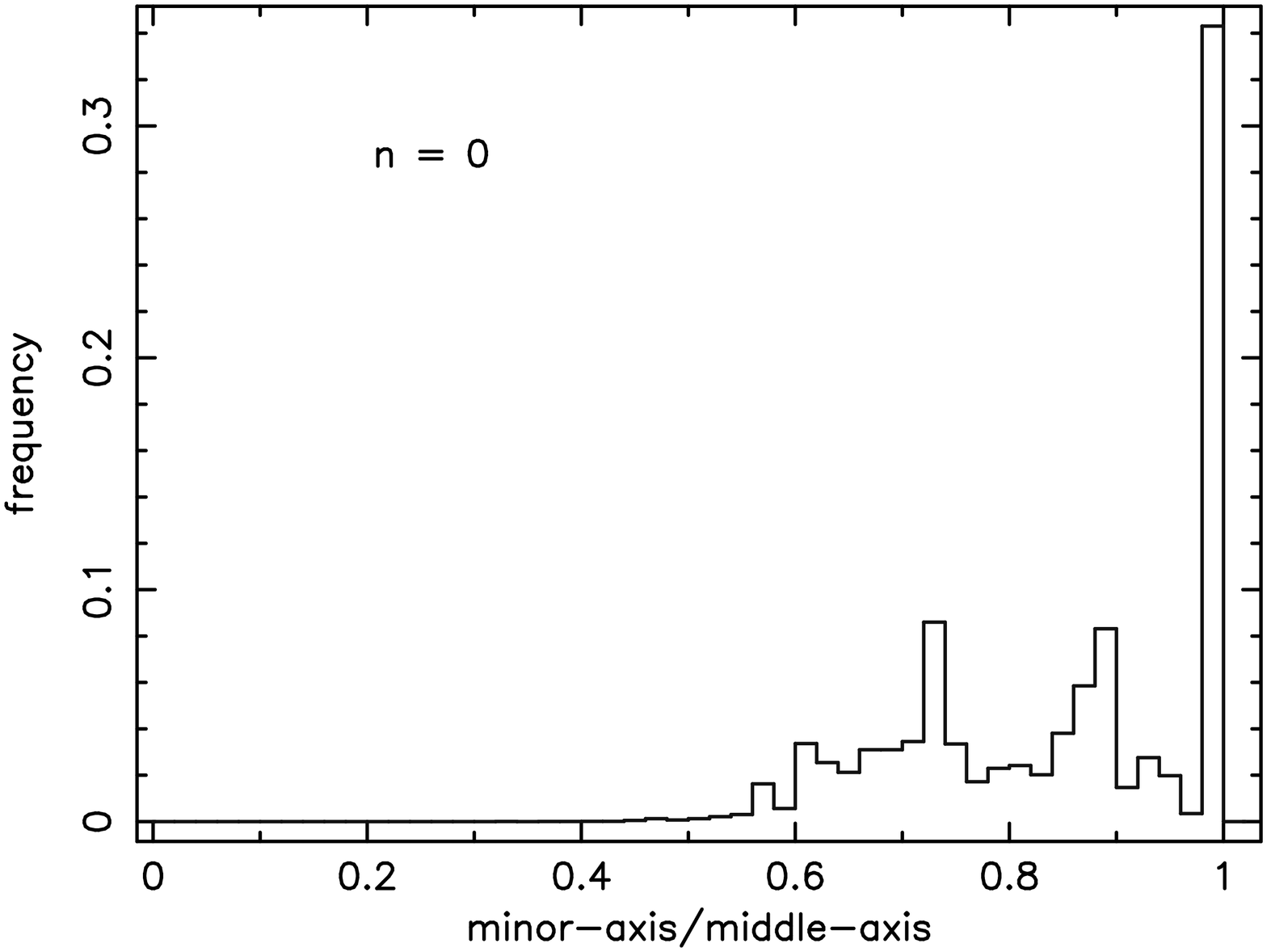,height=7.5cm,width=7.5cm,angle=270} &
\psfig{figure=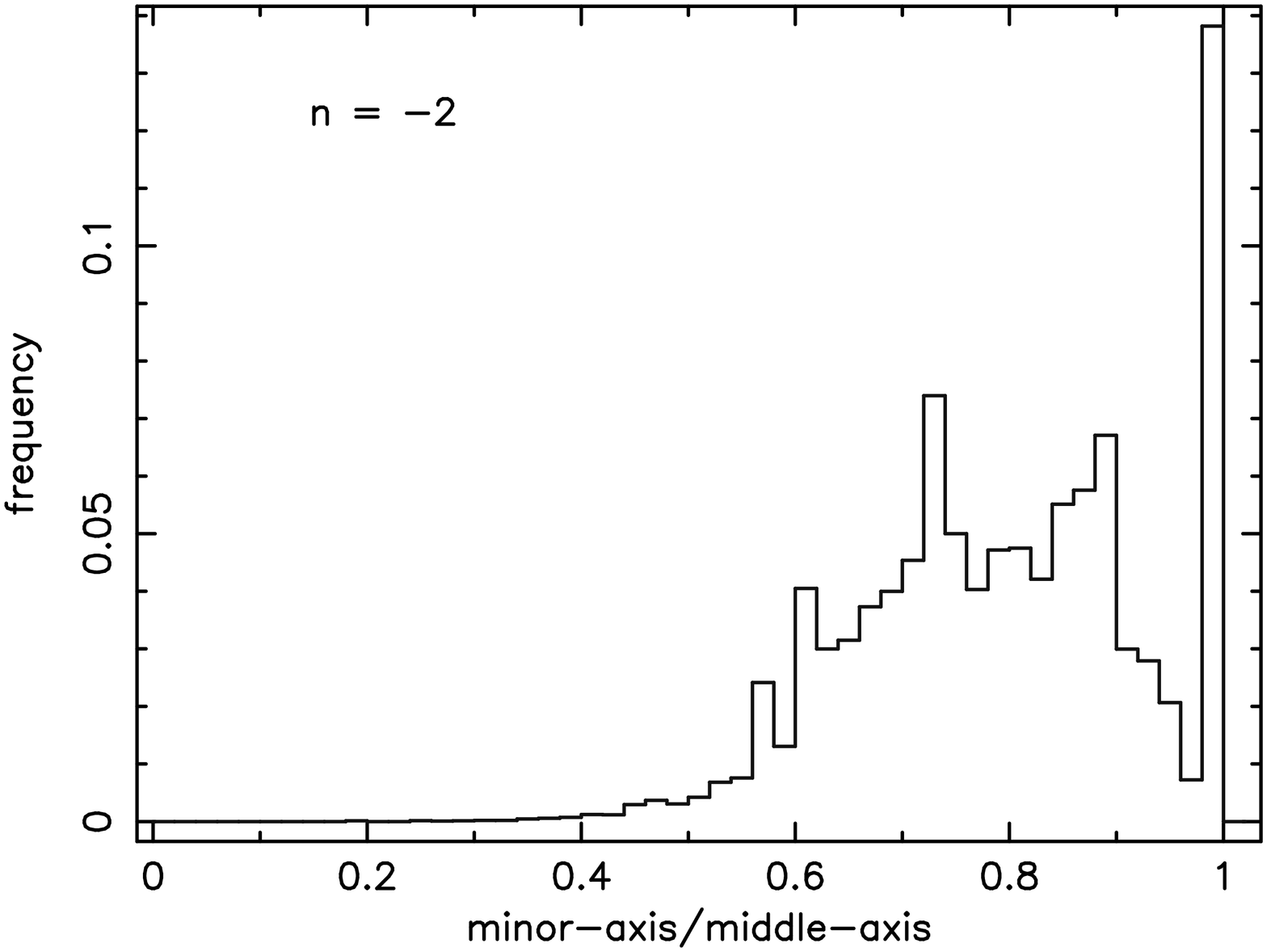,height=7.5cm,width=7.5cm,angle=270} \\
\end{tabular}
\caption{Distribution of axial ratios for halos taken from one $L=128$
realization when half of the box mass is contained in collapsed structures:
$n=0$, mass greater or equal than 8; $n=-2$, mass greater than 50. The upper
panel shows the distribution of triaxalities in the prolateness-ellipticity
plane (with the filled triangle corresponding to the \bm).}
\label{fig:axes}
\end{figure*}

\Fig~\ref{fig:axes} shows the distribution of axial ratios for all
halos of mass greater than or equal to 8 for $n=0$ and greater than
50 for $n=-2$.
The overall observation is that there is no much difference of halo
shapes if one compare realizations of both spectra. In both, the halos
show a wide range of triaxality ranging from prolate to oblate (while
in the \bm\ they are systematically prolate).

\begin{figure}
\centering
\pssilent
\psfig{figure=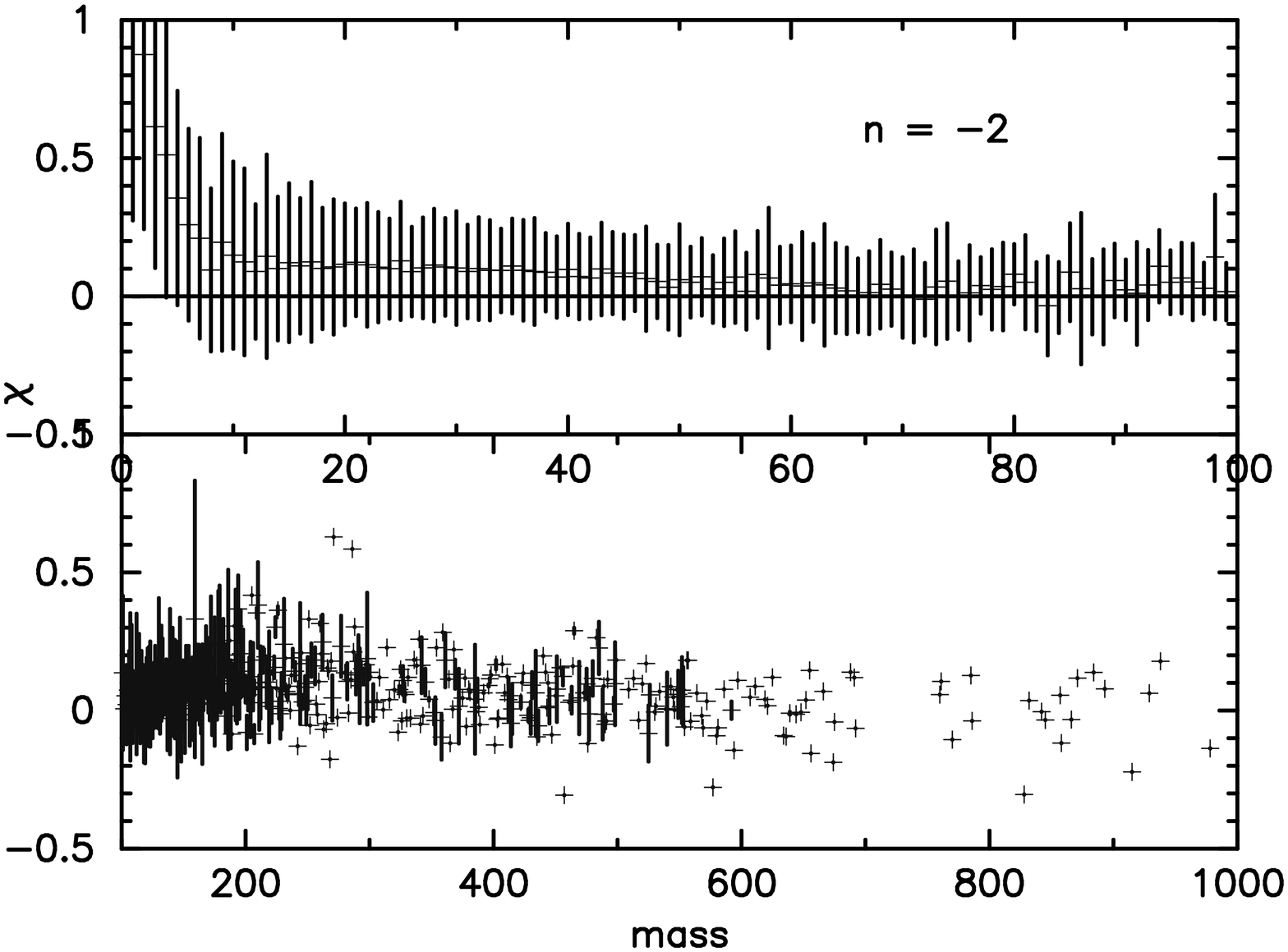,height=6cm,width=8cm,angle=270}
\vspace{0.5cm}
\psfig{figure=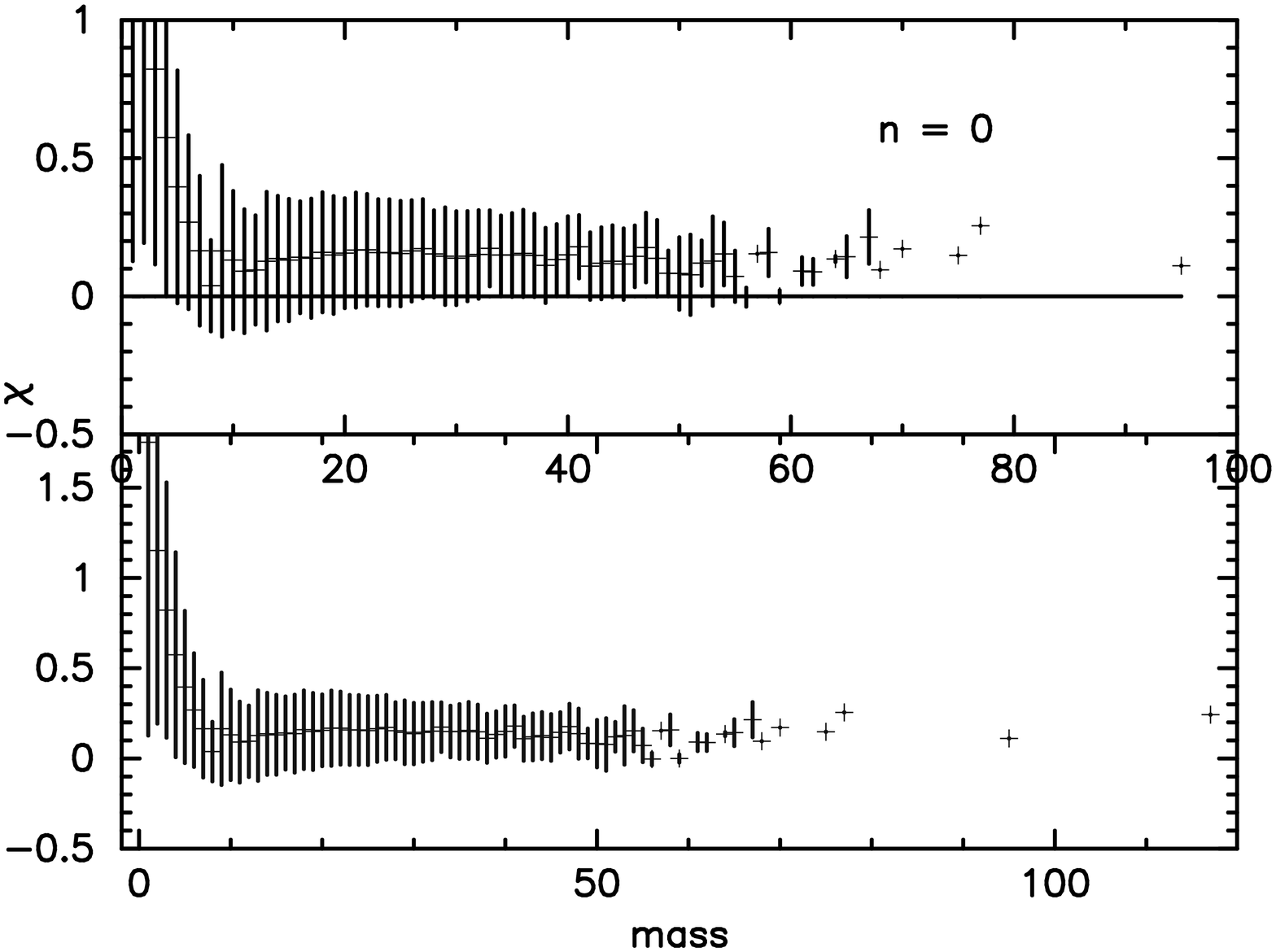,height=6cm,width=8cm,angle=270}
\caption{The relative difference between the assigned and the true
overdensity of halos, for the $L=128$ box at the time when half of its
mass is contained in collapsed regions.}
\label{fig:dmas}
\end{figure}

A drawback of our method comes from the fact that the overdensity
of a collapsing block, $\deltaa$, is not necessarily equal to the mean
overdensity of the resulting halo, $\deltat$.  It is the former value
which we must associate with the halo if the topology of the merger
tree is to be preserved (or at least we must maintain the same
ordering of densities for halos as their parent blocks). The
differences can be quantified in terms of the ratio
$\chi=(\deltaa-\deltat)/\deltaa$ which is
plotted in \Fig~\ref{fig:dmas} at a time when half the mass is in
collapsed structures: we show the mean value plus one sigma error
bars.

Note first that halos of fewer than eight cells have overdensities
which are much less than the assigned one.  These structures are,
however, leftovers of the merging process (the smallest blocks have a
mass of 8 units) and so they should not be considered as collapsed
halos, but rather clouds of intergalactic material to be accreted
later by a neighbouring halo.

For halos of mass 8 or larger the agreement is much better, but
nevertheless the true overdensity of a halo remains systematically
lower than the assigned one.  The effect is largest for $n=0$ where
the mean value of $\chi$ is about 0.15.  For $n=-2$, it varies from
approximately zero in the largest halos to 0.1 in the low mass ones.
The reason for the offset is that high-density cells can contribute to
the overdensity of more than one block.  Refering again to
\Fig~\ref{fig:halo}c, if the region of overlap between the two blocks
were of higher density than its surroundings then the density of the
whole halo would be lower than that of either block from which it is
constructed.  If desired the assigned halo densities could be
systematically reduced to bring them into agreement with the measured
ones; equivalently one could raise the value of the critical density,
$\delta_c$, required for collapse above that of the top-hat model.

More serious is variance of $\chi$, approximately 0.2, which means
that two halos with the same assigned density can have quite disparate
true overdensities.  The model supposes that they collapse at the same
time, whereas the full non-linear evolution would presumably show
otherwise.
We occasionally find some high-mass halos (mass greater than 8) with
big $\chi$, which contribute significantly to the
enlargement of the error bars at those scales. These are effectively
leftovers of the merging process and should not be treated as
collapsed halos in subsequent applications of the method (that is, in a
realistic galaxy formation modelling they should be considered as
sources of material to be accreted at a later stage of the hierarchy).

The variance in $\chi$ is unwelcome but is only one contribution
to the dispersion between overdensity and collapse time.
We note that N-body simulations show for each particle
a poor correspondence between the expected mass of its parent halo
(predicted from the initial conditions) and the true value measured from a
numerical simulation, evolved from the same initial conditions
(White 1995, Bond \etal\ 1992). Moreover gravitational collapse is clearly
not as simple as the spherical model assumes. There is, for example,
no guarantee that underdense regions never collapse or that high-dense
regions will do so (Bertschinger \& Jain 1994).
However, if a simple semi-analytical model of the gravitational clustering
is desired, then the simple relation between collapse
redshift and initial overdensity given by the spherical model seems the
most obvious choice.

\subsection{The number density of high-mass halos}
\label{ssechighm}

Our method passes the test for self-similarity, yet for $n=0$ it
predicts far more high-mass halos than Press-Schechter or other
methods based on similar ideas, such as the \bm.  This is an expected
outcome of our method and points more to a deficiency in
the PS model than anything else, as we attempt to explain below.

Press-Schechter does not count the number of halos of a given mass.
Rather, it counts the fraction of the Universe where, if one were to
put down a top-hat filter of the appropriate mass, the overdensity
would exceed a certain critical threshold.  Regions which just poke
above this threshold for a single position of their centres,
contribute nothing to the mass-function.  It is easy to see that this
becomes increasingly likely as one moves to rarer and rarer objects
(of higher and higher mass).  For these it is much better simply to
count the number of peaks which exceed the threshold density after
filtering on the appropriate scale (on the other hand Peaks Theory
predicts far too many low-mass objects as it does not distinguish
between overlapping halos).  The necessary theory has been
exhaustively analyzed by Bardeen \etal (1986) who showed that
uncorrected PS (without the extra factor of two) underestimates the
number of high-mass halos by a factor $\alpha^{3/2}\nu^3$, where
$\delta_c=\nu\sigma(m)$ (this result is for a gaussian filter but
similar results will hold for all filters with just a small difference
in scaling).  One way to visualize this result is to think of each
peak as having an overdensity profile
\begin{equation}
\nu\approx\nu_0\left(1-{1\over2}\left(r\over R\right)^2\right)
\end{equation}
where $R$ is the radius of the top-hat filter.  It is then easy to
estimate the contribution to the PS mass fraction and to integrate over
all values of $\nu_0$ greater than the threshold, $\delta_c/\sigma$, to
get the total number of halos.  This method suggests that PS should
predict $\sqrt{2/9\pi}\,\nu^3$ times as many halos as Peaks Theory, in
rough agreement with the above for $n\approx -1$ to 0.

A more direct demonstration of the above difference between
Press-Schechter and the actual number of high-mass peaks in the
density field is shown by the numbers in Table~\ref{tab:npeaks}.
Columns 2--9 show the measured number of blocks which exceed the
density threshold given in the first column in each of the eight
sub-grids of mass 512.
\begin{table*}
\begin{minipage}{140mm}
\centering
\caption{Number of blocks of mass 512 ($L=128$)
above the density thresholds $3\sigma$, $2\sigma$, $\sigma$: (a)
 $n=-2$, (b) $n=0$.  Columns 2--9 correspond to each of the eight
 sub-grids used for that level of refinement, Column~10 shows the
number of isolated peaks and Column~11 the PS prediction of the
expected number of halos of this mass.}
\label{tab:npeaks}
\begin{tabular}{@{}rrrrrrrrrcc@{}} \hline
	(a) XYZ & single &   &   &   &   &   &   &   &combined&PS expected \\
	   grid& grid & X & Y & Z & XY & XZ & YZ & XYZ & grids & number\\
 \hline
	$\geq 3\sigma$ &5&7&4&1&6&7&6&8&18& 5.5$\pm$2.3 \\
	$\geq 2\sigma$ &88&87&92&77&92&87&79&88&109& 93.2$\pm$9.7 \\
	$\geq 1\sigma$ &628&654&630&662&628&648&639&647&281& 650$\pm$26 \\
\end{tabular}
\vspace{0.5cm}
\begin{tabular}{@{}rrrrrrrrrcc@{}} \hline
	(b) XYZ & single &   &   &   &   &   &   &   &combined&PS expected \\
	   grid& grid & X & Y & Z & XY & XZ & YZ & XYZ & grids & number\\
 \hline
	$\geq 3\sigma$ &6&7&4&9&4&7&8&7&38&5.5$\pm$2.3 \\
	$\geq 2\sigma$ &99&99&96&82&90&86&99&90&324&93.2$\pm$9.7 \\
	$\geq 1\sigma$ &630&641&649&656&652&617&670&677&671&650$\pm$26 \\
\end{tabular}
\end{minipage}
\end{table*}
These agree with the Press-Schechter prediction, as indeed they should
by construction.  When we combine the various grids, however, an
interesting thing happens.  Column~10 shows the number of separate,
(\ie non-overlapping), overdense blocks in the combined grid.  At high
overdensity all the halos we have identified are distinct (they exceed
the threshold for just one position of the smoothing grid).  The total
number of halos is therefore greatly in excess of the PS prediction
and far closer to that given by Peaks Theory.  For $n=-2$ the excess
is approximately a factor of three which brings them into agreement
once the PS prediction has
the extra factor of two applied.  For $n=0$, however, the difference
is much larger and the number of peaks is a factor of 3-4
larger than even the corrected PS estimate.  This goes in some way to
explaining the difference between the PS prediction and the measured
cumulative mass function in \Fig~\ref{fig:cummf}.

At lower overdensity the disagreement is much less severe. One should
note that for $n=-2$ there is a gross underestimate of $1\sigma$ peaks
compared to the values obtained in each sub-grid. This is simply a
consequence of the Peaks methodology. Remember that for each sub-grid
we are just measuring the fraction of the total number of blocks above
the threshold, while in the case of combined grids we simply count the
number of peaks.  Because for $n=-2$ the peaks are larger and more
clustered, there is a great chance of finding blocks sitting next each
other which are above the imposed threshold. Consequently, if we are
only selecting the peaks many of those blocks will be discarded. This
situation does not arise for $n=0$, where the peaks are smaller and
more evenly distributed.

Our use of overlapping grids is therefore crucial.  They ensure that
all halos are approximately centred within one of the grid cells.
Other methods, such as the \bm, which have fixed borders between mass
cells, have difficulty in detecting structures that cross cell
boundaries and are, by construction, forced to agree with
Press-Schechter.  This can lead to a gross underestimate of the number
of rare, high-mass peaks, especially for steep spectra.
We are not saying that our method necessarily gives a better
description of the growth of structure in the Universe because all
these theoretical models are highly idealized.  Substructure may
lengthen collapse times and tidal field may need to be taken into
account.  Nevertheless, given the simplified prescription which we
have adopted, our method does at least seem to detect the correct
number of high-mass halos, and many more than other methods.

\section{Discussion}
\label{secdiscuss}

We have presented a new method of constructing a hierarchical merger
tree based on actual realizations of the linear density field.  We
smooth on a set of interlaced, cubical grids on a variety of mass
scales, then order in decreasing density.  We run down the resulting
list, merging together overlapping blocks to form collapsed halos.
The main properties of our model are as follows:
\begin{itemize}
\item The model exhibits the scaling behaviour expected from power-law
spectra.
\item For a flat, white-noise spectrum, $n=0$, the mass function is
well-fit by the PS model, provided that we raise the normalization by
a factor of 3-4.  This difference arises because of a deficiency
in the PS method which fails to count the correct number of massive,
rare (high-$\nu$) peaks.  The dynamic range for the masses of halos
for this spectrum is quite small---at a time when half the box is in
collapsed structures, the mass of the largest halo is just 125 cells,
even for the box of side $L=256$.
\item The mass function for a steeper spectrum, $n=-2$, lie much
closer to the PS prediction (with the usual factor of two increase in
normalization), but show small kinks at masses of 64, 512 and 4096,
corresponding the the masses of smoothing blocks---there is a slight
deficit of halos of smaller, and an excess of halos at larger, mass.
\item The collapsed halos tend to be triaxial with a wide mixture
of prolateness and oblateness (contrary to the \bm).
\item The mean overdensity of halos tends to be lower than that
assigned to them by our scheme, and to have a scatter of about 20
percent about the mean value.  While undesirable, since it can reverse
the collapse ordering in the tree, this can be partialy
cured in subsequent applications of the method.
\end{itemize}

A shortcoming of our method is that the minimum halo mass is eight
cells with a corresponding loss of dynamical range. We see in
\Fig~\ref{fig:cummf} that for $n=-2$ we achieve approximately a
mass-range of approximately two and a half decades when half the mass
is contained in collapsed objects, while for $n=0$ we get only
slightly more than one decade. This is because $n=-2$ corresponds to
a flat spectrum, $\sigma\propto M^{-1/6}$, with almost all scales
collapsing simultaneously, while for $n=0$ the spectrum decreases more
rapidly, $\sigma\propto M^{-1/2}$, and the peaks are much more
isolated.  Fortunately, the physically-motivated CDM spectrum can be
fitted by an $n=-2$ spectrum over a significant mass range, and so the
loss of dynamical range is not so important in a realistic
application of the method.  All these mass-ranges can be extended if
we allow a larger fraction of the box to collapse (as we must if we
wish to model a cluster of galaxies, for example), but at the risk of
losing a fair representation of the power spectrum on scales
approaching the box-size.

If we simulate a large portion of the Universe, of mass say
$10^{16}\Msun$, in a box of $L=128$, then a block of 8 cells
corresponds to $3.8\times 10^{10}$\Msun, which could represent at
best a dwarf galactic halo.  If the box represents a large galactic
halo of mass $10^{13}\Msun$, then we can resolve down almost to
globular cluster scales.

\begin{figure}
 \centering
 \pssilent
 \psfig{figure=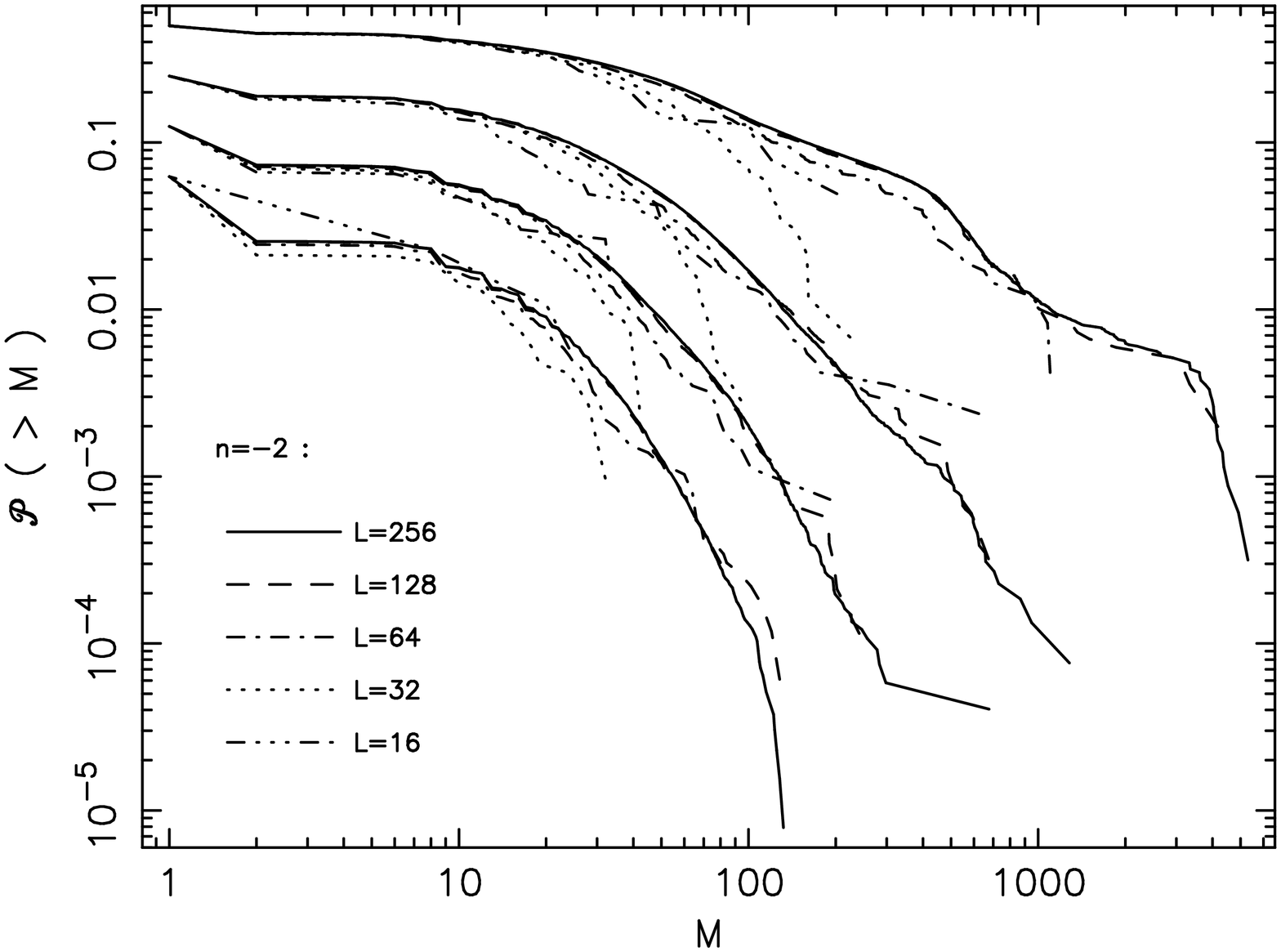,height=6cm,width=8cm,angle=270}
 \psfig{figure=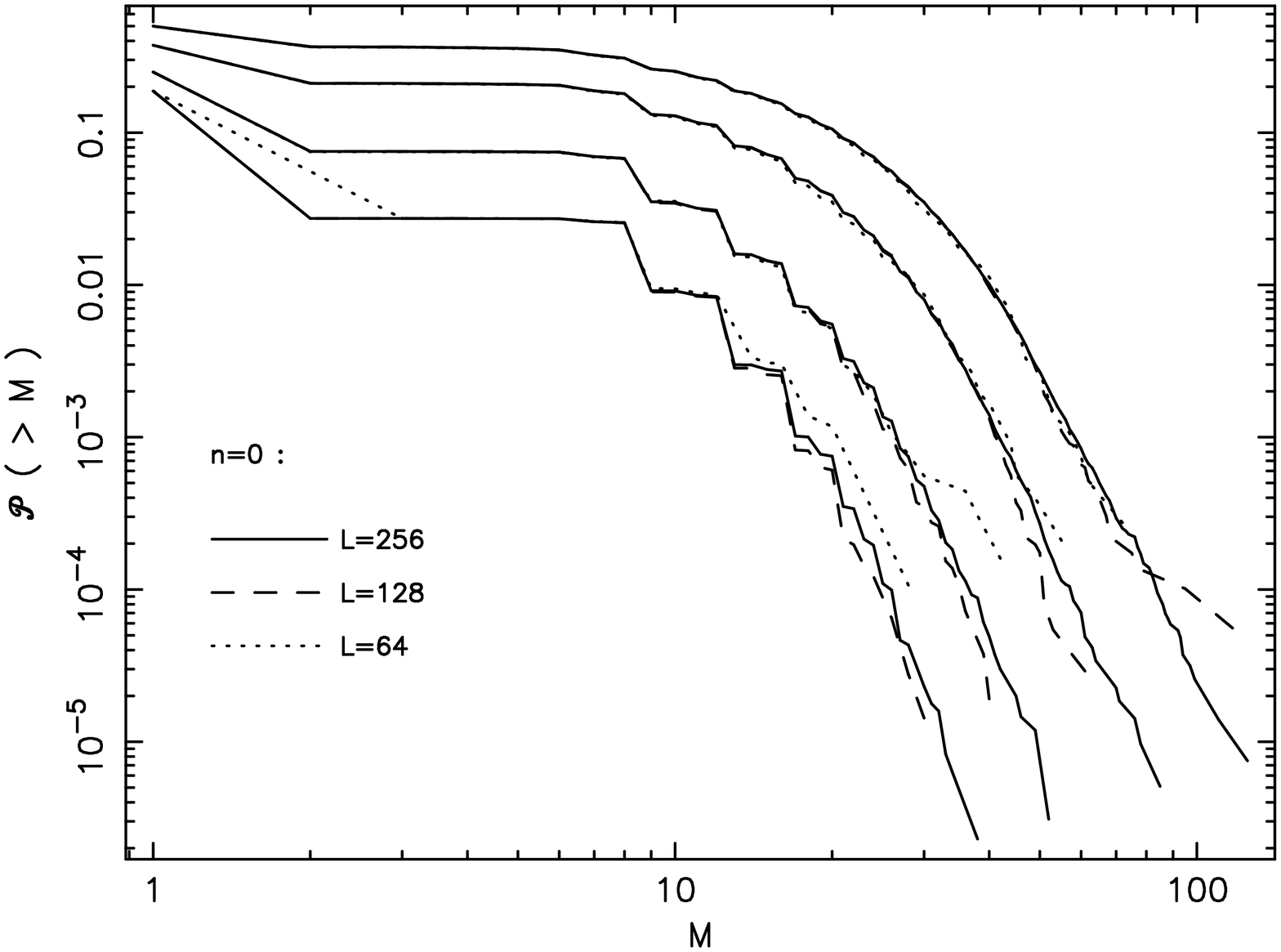,height=6cm,width=8cm,angle=270}
 \caption{The cumulative mass-function for boxes of
 variable resolution, as indicated: (a) $n=-2$, (b) $n=0$.  The collapsed
 fractions are the same as those corresponding to \Fig~\ref{fig:cummf}
 respectively.}
 \label{figvarl}
\end{figure}

We also have carried out simulations with a range of box-lengths, $L=16$,
32, 64, 128, and 256, in order to show the effect of variable resolution (we
were not able to perform the $L=16,32$ simulations for $n=0$ due to the
lack of dynamical range). The base-cells in each case correspond to one
set of blocks of side $256/L$ in the $L=256$ simulation. The results
are presented in Fig.~\ref{figvarl} which shows the cumulative mass functions
sampled at the same collapsed fractions of the box as those corresponding
to Fig.~\ref{fig:cummf}. The shape of the mass spectrum is similar but
with a slight increase in dynamic range as one moves from $L=64$ to $L=256$.

Whether our method provides a better description of the formation of
structure than other methods remains to be seen.  The \bm\ in
particular seems to give good agreement with N-body simulations (Lacey
\& Cole 1994) which themselves are approximately fit by modified
Press-Schechter models (\eg Gelb \& Bertschinger 1994).  However,
there is a limited dynamical range in the simulations, the results are
sensitive to the precise model for identifying collapsed halos and the
critical overdensity for collapse is usually taken as a free
parameter.  Given these caveats it is hard to tell whether the fits
are mediocre, adequate or good.

One advantage of our scheme is that it is based on an actual
realization of a density field which can be used as the starting point
for an N-body simulation.  Thus we will not be limited to a
statistical analysis, but will be able to directly compare individual
structures identified in the linear density field with non-linear
halos that form in the simulation.  Initial results from other studies
(Bond \etal 1992, Thomas \& Couchman 1992 ) suggest that the
correspondence is approximate at best, and we may be forced to
consider the effect of tidal fields on a halos evolution.  We have not
yet carried out the necessary N-body simulations because we have not
up to now had access to the necessary super-computing facilities to
evolve (and analyze!) a $256^3$ box of particles.  Such datasets will
soon become available as part of the Virgo Consortium project on the
UK's Cray T3D facility and we intend to report the results in a
subsequent paper.

Nevertheless, even in the absence of the numerical tests, we feel that
our method is a viable alternative to other methods of calculating the
merging history of galactic halos.  It passes the test of
self-similarity yet predicts more high-mass halos than other methods.
It has the disadvantage of losing a factor of eight in resolution at
low-masses, but above this it has a smooth mass-spectrum and is not
restricted to masses which are a power of two.
We intend to contrast
the predictions of the \bm\ and this current method in models of
galaxy formation such as those discussed by Kauffmann, White \&
Guiderdoni (1993) and Cole \etal (1994), and explore the role of pre-galactic
cooling flows (Nulsen \& Fabian 1995).

\section*{Acknowledgments}
DDCR would like to acknowledge support from JNICT (Portugal) through
program PRAXIS XXI (grant number BD/2802/93-RM).  Part of this paper
was written while PAT was at the Institute for Theoretical Physics at
Santa Barbara and as such was supported in part by the National
Science Foundation under Grant Number PHY89-04035.  The paper was
completed while PAT was holding a Nuffield Foundation Science Research
Lectureship.  We would like to thank Shaun Cole for providing us with
a copy of the Block Model program.  The production of this paper was
aided by use of the STARLINK Minor Node at Sussex.

\end{document}